\documentclass[floatfix,prd,a4paper,eqsecnum,nofootinbib,preprint]{revtex4}
\usepackage{epsfig}
\usepackage{graphicx}
\usepackage{bm}

\def\bk{{\bm{k}}}
\def\bq{{\bm{q}}}
\def\bp{{\bm{p}}}
\def\bn{{\bm{n}}}
\def\bN{{\bm{N}}}
\def\bmm{{\bm{m}}}

\def\muhat{\hat{\bm{\mu}}}
\def\Tr{\text{Tr}\,}

\def\e{{\text{e}}}
\def\o{{\text{o}}}

\def\Eq#1{Eq.~(\ref{#1})}

\renewcommand\Im{\text{Im}\,}
\renewcommand\Re{\text{Re}\,}

\begin{document}
\title{Order from disorder in lattice QCD at high density}
\author{Barak Bringoltz}
\affiliation{School of Physics and Astronomy, Raymond and Beverly
Sackler Faculty of Exact Sciences, Tel Aviv University, 69978 Tel
Aviv, Israel}
\begin{abstract}
We investigate the
properties of the ground state 
of strong coupling lattice QCD at finite density. 
Our starting point is the effective Hamiltonian for color singlet objects,
which looks at lowest order as an antiferromagnet, and describes meson physics with a fixed baryon number background.
We concentrate on uniform baryon number backgrounds (with the same baryon number on all sites), for which the ground state was extracted in an earlier work, and calculate the
dispersion relations of the excitations. Two types of Goldstone boson
emerge.
The first, antiferromagnetic spin waves, obey a linear dispersion relation.
The others, ferromagnetic magnons, have energies that are
quadratic in their momentum. 
These energies emerge only when fluctuations around the large-$N_c$ ground
state are taken into account, along the lines of ``order from disorder''
in frustrated magnetic systems.
Unlike other spectrum calculations in order from disorder, we employ the Euclidean path integral.
For comparison, we also present a Hamiltonian calculation using
a generalized Holstein--Primakoff transformation.
The latter can only be constructed for a subset of the cases we
consider.

\end{abstract}
 \maketitle

\section{\label{sec:intro}Introduction}

The study of quantum chromodynamics at high density is of fundamental importance.
In recent years the field has attracted wide attention with 
renewed interest in color superconductivity (CSC). For a review see \cite{Rajagopal:2000wf}.

CSC at high density is a prediction of weak-coupling analysis. It is
imperative to confirm this prediction by non-perturbative
methods. Bringoltz and Svetitsky \cite{Bringoltz:2002qc} studied high-density
quark matter with strong-coupling lattice QCD (see also \cite{strong_coupling}). This theory is
described by an effective Hamiltonian for color singlet objects. At lowest order in the inverse coupling, the Hamiltonian looks like an antiferromagnet, and describes the dynamics of meson-like objects within a fixed background of baryon number. Higher orders in the inverse coupling add the effective Hamiltonian with kinetic terms for the baryons that are very hard to treat \cite{Bringoltz:2002qc}. The global symmetry group of the antiferromagnet depends on the
formulation of the lattice fermions. For naive, nearest-neighbor
fermions, studied here and in Ref.~\onlinecite{Bringoltz:2002qc}, the
symmetry is $U(4N_f)$. The representation of $U(4N_f)$ carried by the
fields on each site depends on the baryon number on that site.

From the antiferromagnetic effective Hamiltonian one can derive a path integral for a
nonlinear sigma model (NLSM). The path integral can be studied in the
large-$N_c$ limit with semiclassical methods. 

Following Ref.~\onlinecite{Bringoltz:2002qc}, we choose to work with baryon number backgrounds that have the same baryon number $B$ on all sites with $|B|=1,2,\dots,2N_f-1$. These backgrounds are close to lattice saturation (that happens for $B=2N_f$), but involve relatively simple calculations; The sigma fields on all sites belong to the same manifold and represent the same number of independent fluctuations (see \cite{Bringoltz:2002qc}). Any other distribution of baryon number can also be treated with the NLSM, but will involve sigma fields whose structure changes from site to site. This makes the calculations of the semi-classical ground state and excitations less straight forward than presented here and in Ref.~\onlinecite{Bringoltz:2002qc}. We leave the treatment of non-uniform baryon number backgrounds, together with the kinetic terms for the baryons to future study, as a step towards comparing our results to those of grand canonical approaches.

For the baryon number background we study, the
classical ($N_c=\infty$) vacuum of the NLSM is hugely degenerate: If
one aligns the $\sigma$ fields on the even sites,
the field on each odd site can independently wander  a
submanifold of the original $\sigma$ manifold. It was shown in
\cite{Bringoltz:2002qc} that fluctuations in $O(1/N_c)$ couple the $\sigma$
fields on the odd sites to each other and give an effective action that lifts the
classical degeneracy. The ground state breaks $U(4N_f)$. 

Here we treat the problem in the context of ``order from
disorder'' \cite{Aharony,Ord_disorder,double_Xchange,Kagome,Sachdev,Henley}, a phenomenon known in
the study of  condensed matter systems. Its essence is that fluctuations of
quantum, thermal or even quenched nature can lift a classical
degeneracy. A famous example is the Kagom{\' e}
antiferromagnet \cite{Kagome} whose classical ground state energy is invariant
under correlated rotations of local groups of spins, leading to a
degeneracy exponential in the volume. This system has modes with zero
energy for all momentum. These zero modes obtain nonzero energy due to quantum
fluctuations. The same thing happens in our NLSM. Here we identify the zero modes, show how
they get nonzero energy of $O(1/N_c)$ and calculate their dispersion relations.

We end up with two kinds of Goldstone boson. The first are
 antiferromagnetic spin waves with a linear dispersion
relation. The second kind are ferromagnetic
magnons with a quadratic dispersion relation. This
is consistent with the loss of Lorentz invariance due to finite
density. 
This set of excitations falls into the pattern described by Chadha and Nielsen \cite{Nielsen:hm} and by Leutwyler 
\cite{Leutwyler:1993gf} in their
studies of nonrelativistic field theories. In general, a nonrelativistic system that undergoes spontaneously symmetry breakdown
can posses two types of Goldstone boson. The energy of type I bosons is an odd power of their momentum.
 Their effective field theory has a second time derivative which means that
each excitation is described by one real field as in the case of a relativistic scalar field. 
Type II bosons have a dispersion relation 
that contains an even power of the momentum. They can appear only in nonrelativistic theories having properties similar to a Schr\"{o}dinger field theory, namely theories whose action has only a first time 
derivative. As in the Schr\"{o}dinger case, each excitation is
described by a complex field, or two real fields. 
The number of Goldstone fields is of course equal to the number $n_G$ of broken generators. The counting of massless excitations is summarized by the Chadha--Nielsen counting rule, 
\begin{equation}
\label{count}
n_{I}+2n_{II} \geq n_G,
\end{equation}
Where $n_I(n_{II})$ are the number of type I and II Goldstone bosons. A well known example 
is the $SU(2)$ spin system with a collinear ground state
, where $n_G=2$. The antiferromagnet has two spin
waves with a linear dispersion relation ($n_I=2,n_{II}=0$) while the
ferromagnet has one magnon with a quadratic dispersion relation
($n_I=0,n_{II}=1$). As mentioned above in our case both $n_I$ and $n_{II}$ are nonzero.

Most discussions of ``order from disorder'' work in a Hamiltonian (but see \cite{Sachdev,Henley}).
We employ instead the Euclidean path integral. In Appendix~\ref{app:Hamiltonian} we also present a 
Hamiltonian calculation using
a generalized Holstein-Primakoff \cite{Randjbar-Daemi:1992zj} transformation. The latter can only be constructed for a subset of the
cases we consider. The Euclidean calculations are less restrictive in their applicability.

\section{The Nonlinear Sigma model\label{sec:NLSM}}

In Ref.~\onlinecite{Bringoltz:2002qc} the problem of
lattice QCD at finite densities was transformed to a nonlinear sigma model. In this section we review this approach in order to make the current work self-contained.  
The starting point is strong coupling QCD with naive fermions. This theory is
described by an antiferromagnetic (AF) effective Hamiltonian, which is
invariant under a global $U(N)$ symmetry with $N=4N_f$. This quantum Hamiltonian is given by
\begin{equation}
\label{Hqm}
H_{\text{eff}}=\frac{J}2 \sum_{\bn \muhat} Q_{\alpha \beta}(\bn)
Q_{\beta \alpha}(\bn+\muhat).
\end{equation}
Here $\alpha$ and $\beta$ are Dirac-flavor indices taking values from
$1$ to $N$, and $Q_{\alpha \beta}(\bn)$ are
the generators of $U(N)$ on site $\bn$. The Hilbert space on each site forms an irreducible
representation of $U(N)$ which corresponds to a rectangular Young
tableau with $N_c$ columns. The number $m$ of rows in the Young
tableau can vary from site to site and
is determined by the local baryon number $B_\bn$ according to
\begin{equation}
\label{B}
m_\bn=B_\bn+N/2.
\end{equation}
Therefore, a first step in this Hamiltonian approach is to fix the local baryon number. This restricts the diagonalization of~(\ref{Hqm}) to a certain sector of the Hilbert space, in contrast to other approaches that fix a chemical potential \cite{Rajagopal:2000wf,strong_coupling} and let the local baryon number fluctuate. This point makes the comparison between the two approaches nontrivial and out of the scope of this paper. It also reduces the possibility of sketching a phase diagram to identifying the ground state for a given configuration of $B_\bn$, as was done in Ref.~\onlinecite{Bringoltz:2002qc}.

To write the Euclidean path integral we use a generalized spin-coherent-state
representation, which turns out to be especially useful in the large-$N_c$
limit. The result is a nonlinear sigma model (NLSM), which we proceed to describe.

\subsection{Elementary fields}
Following \cite{Bringoltz:2002qc} we study only the case of uniform baryon
density, $B_\bn=B>0$. The $\sigma$ field at site $\bn$ is an $N\times N$
hermitian, unitary matrix that represents the coset space
$U(N)/[U(m)\times U(N-m)]$. It can be written as a unitary rotation
of the standard matrix $\Lambda$,
\begin{equation}
\sigma_\bn=U_\bn \Lambda U^{\dag}_\bn \label{sigma_n},
\end{equation}
with
\begin{equation}
\Lambda = \left( \begin{array}{cc}
        1_{m}     & 0 \\
        0 &     -1_{N-m}        \end{array}     \right). \label{Lambda}
\end{equation}
A convenient
parameterization \cite{Randjbar-Daemi:1992zj} for the unitary matrix $U_\bn$
is\footnote{This can be related to the form used in \cite{Bringoltz:2002qc}
  by setting $\phi=a\sin{\sqrt{a^{\dag}a}}/\sqrt{a^{\dag}a}$. The main
  advantage of this $\phi$ parameterization is that the
kinetic part of the action is quadratic in $\phi$
[see~(\ref{SeffX})].} 
\begin{equation}
\label{U_phi}
U_\bn=\left( \begin{array}{ccc}
        \sqrt{1-\phi_\bn \phi^{\dag}_\bn} & & \phi_\bn      \vspace{0.3cm}  \\
        -\phi^{\dag}_\bn & & \sqrt{1-\phi^{\dag}_\bn \phi_\bn} \end{array} \right).\end{equation}
Here $\phi_\bn$ is an $m \times (N-m)$ complex matrix that furnishes
coordinates for $U(N)/[U(m) \times U(N-m)]$. 
Inserting into \Eq{sigma_n}, we obtain
\begin{equation}
\label{sigma_phi}
\sigma_\bn=\left( \begin{array}{ccc}
        1-2\phi_\bn \phi^{\dag}_\bn & & -2\phi\sqrt{1-\phi^{\dag}_\bn
          \phi_\bn}  \vspace{0.3cm} \\
        -2\sqrt{1-\phi^{\dag}_\bn \phi_\bn} \phi^{\dag}_\bn & & -1 +
        2\phi^{\dag}_\bn \phi_\bn  \end{array} \right).
\end{equation}
Note that $N_c$ does not enter the definition of these degrees of
freedom.

\subsection{The action and the ground state}

The action of the NLSM is
\begin{equation}
\label{S}
S=\frac{N_c}2\int d\tau\left[-\sum_{\bn}
\Tr\Lambda U^{\dag}_{\bn} \partial_\tau U_{\bn} +\frac
J2\sum_{\bn\mu} \Tr \sigma_\bn \sigma_{\bn+\muhat}\right].
\end{equation}
The overall $N_c$ factor allows a systematic
treatment in orders of $1/N_c$.  The
classical ground state is found by minimizing the
action, which gives field
configurations that are $\tau$ independent and  minimize the
interaction. For $B\neq 0$, this leaves a huge manifold of
degenerate  field configurations. 

To clarify this important
point, consider the energy of one link,
\begin{equation}
\label{site2}
E=\frac{J}2 \Tr \sigma_1 \sigma_2.
\end{equation}
To minimize $E$ we work in a basis where $\sigma_1=\Lambda$. The
analysis in \cite{Bringoltz:2002qc} then shows that $\sigma_2$ can wander freely
in the manifold $U(m)/[U(2m-N)\times U(N-m)]$, a submanifold of
$U(N)/[U(m)\times U(N-m)]$. A ground state of the infinite lattice can
be constructed replicating $\sigma_1$ and $\sigma_2$ on the even and
odd sites of the lattice. Thus all the $\sigma$ fields on the even sites ``point'' to
$\Lambda$ while on the odd sites, each $\sigma$ field wanders
independently in the submanifold. This classical ground state has a huge degeneracy,
exponential in the volume. We proceed to analyze its stability.

In \cite{Bringoltz:2002qc} it was shown that $O(1/N_c)$ fluctuations lift the
degeneracy and choose a collinear ground state, where all the $\sigma$
fields on the odd sites point in the same direction.
Without loss of generality the ground state can be chosen to be
\begin{eqnarray}
\sigma_{\text{even}}&=&\Lambda,     \nonumber       \\
\sigma_{\text{odd}}&=&\Lambda_2 \equiv \left( \begin{array}{ccc}
        1_{2m-N} & 0 & 0 \\
        0 & -1_{N-m} & 0 \\
        0 & 0 & 1_{N-m}  \end{array} \right)=V\Lambda V^{\dag}, \label{gs}
\end{eqnarray}
with 
\begin{equation}
\label{V}
V=\left( \begin{array}{ccc}
        1 & 0 & 0       \\
        0 & 0 & 1       \\
        0 & -1 & 0 \end{array}  \right).
\end{equation}
This ground state breaks
 $U(N)$ to $U(2m-N)\times U(N-m) \times U(N-m)$,
with $2(3m-N)(N-m)$ broken generators. This was the main result
of Ref.~\onlinecite{Bringoltz:2002qc}.

\section{Fluctuations around the ground state \label{sec:fluc}}
In this section we write an effective action for fluctuations around the
 ground state up to $O(1/N_c)$. Working at $O(1)$
first, we identify the zero modes that correspond to the classical
degeneracy. We then show how quantum fluctuations give them nonzero energy.
\subsection{Effective action and $O(1)$ dispersion relations \label{sec:O1}}
The fields $\phi$, suitably shifted, can be identified with the
Goldstone bosons around the ground state~(\ref{gs}). For
the even sites, $\phi=0$ indeed gives $\sigma=\Lambda$. For the odd
sites, the vacuum is at 
$$
\phi=\left( \begin{array}{c}
            0 \\ 1_{N-m} \end{array} \right),
$$
where the upper part of $\phi$ is a $(2m-N)\times (N-m)$ matrix.
We therefore shift $\phi$ on the odd sites according to
\begin{eqnarray}
\label{diff_odd}
U'&\equiv &VU, \nonumber \\
\sigma'(\phi)&\equiv &V\sigma(\phi) V^{\dag}.
\end{eqnarray}
This gives $\sigma'(0)=\Lambda_2$. We drop the primes henceforth.

For later convenience we write $\phi$ as
\begin{equation}
\label{phi}
\phi=\left( \begin{array}{c}    \chi \\ \pi   \end{array} \right).
\end{equation}
Here $\pi$ is an $(N-m)-$~dimensional square matrix while $\chi$ has $(2m-N)$ rows
and $(N-m)$ columns. Both are complex. Thus
\begin{equation}
\label{even}
\sigma_{\text{even}}=\left( \begin{array}{ccc}
        1-2\chi \chi^{\dag} & -2\chi \pi^{\dag} & -2\chi \sqrt{1-\phi^{\dag}\phi}        \vspace{0.3cm} \\
        -2\pi \chi^{\dag} & 1-2\pi \pi^{\dag} & -2\pi \sqrt{1-\phi^{\dag}\phi} \vspace{0.3cm} \\
        -2\sqrt{1-\phi^{\dag}\phi} \chi^{\dag} & -2\sqrt{1-\phi^{\dag}\phi} \pi^{\dag} & -1+2\phi^{\dag} \phi \\      \end{array}     \right),
\end{equation}
and
\begin{equation}
\label{odd}
\sigma_{\text{odd}}=\left( \begin{array}{ccc}
       1-2\chi \chi^{\dag} & -2\chi \sqrt{1-\phi^{\dag}\phi} & 2\chi \pi^{\dag} \vspace{0.3cm} \\
        -2\sqrt{1-\phi^{\dag}\phi} \chi^{\dag} & -1+2\phi^{\dag} \phi & 2\sqrt{1-\phi^{\dag}\phi} \pi^{\dag}  \vspace{0.3cm} \\
        2\pi \chi^{\dag} & 2\pi \sqrt{1-\phi^{\dag}\phi} & 1-2\pi \pi^{\dag} \\ \end{array} \right).
\end{equation}
Each submatrix in Eqs.~(\ref{even})--(\ref{odd}) has dimensions as
indicated in~(\ref{Lambda}).

After scaling $\phi \rightarrow \phi/\sqrt{N_c}$ the
action takes the form,
\begin{eqnarray}
S&=&\Tr \int d\tau\sum_\bn \left[ \chi^{\dag}_\bn \partial_{\tau} \chi_\bn +\pi^{\dag}_\bn \partial_{\tau} \pi_\bn \right] \nonumber \\
&&\qquad +J\sum_{\bn \muhat} \left[ \pi_\bn \pi^{\dag}_\bn +\pi_{\bn+\muhat} \pi^{\dag}_{\bn+\muhat} -\pi_\bn \pi_{\bn+\muhat} -\pi^{\dag}_\bn \pi^{\dag}_{\bn+\muhat} \right] \nonumber \\
&&\qquad +\frac{J}{\sqrt{N_c}} \sum_{\bn \muhat} \left[ \chi^{\dag}_\bn \chi_{\bn+\muhat} \pi_\bn  - \chi^{\dag}_{\bn+\muhat} \chi_\bn \pi_{\bn+\muhat}  +h.c. \right] \nonumber \\
&&\qquad +\frac{J}{N_c} \sum_{\bn \muhat} \left[\chi_\bn \chi^{\dag}_\bn \chi_{\bn+\muhat} \chi^{\dag}_{\bn+\muhat} -\pi_\bn \pi_{\bn+\muhat}^{\dag} \pi_{\bn+\muhat}^{\dag} \pi_{\bn+\muhat} -\pi_{\bn+\muhat} \pi_{\bn+\muhat}^{\dag} \pi_\bn^{\dag} \pi_\bn \right.  \\
&& \qquad \qquad \qquad -\pi_\bn \pi_\bn^{\dag} \chi^{\dag}_{\bn+\muhat} \chi_{\bn+\muhat} -\pi_{\bn+\muhat} \pi_{\bn+\muhat}^{\dag} \chi^{\dag}_\bn \chi_\bn \nonumber \\
&& \qquad \qquad \qquad +\frac12 \left( \pi_{\bn+\muhat} \pi_{\bn+\muhat}^{\dag} \pi_{\bn+\muhat} \pi_\bn +\pi_{\bn+\muhat} \pi_\bn \pi_\bn^{\dag} \pi_\bn \nonumber \right. \\
&& \qquad \qquad \qquad \qquad \left. \left. +\pi_{\bn+\muhat} \chi_{\bn+\muhat}^{\dag}
    \chi_{\bn+\muhat} \pi_\bn + \pi_{\bn+\muhat} \pi_\bn
    \chi^{\dag}_\bn \chi_\bn + h.c. \right) \right]  \nonumber \\ 
&& \qquad +O(1/N_c^{3/2}). \nonumber \label{SeffX}
\end{eqnarray}
The AF structure of the action demands the introduction of an fcc
lattice. We write the fields $\phi_\bn \equiv \phi^A_\bN$ with
$A=(\text{even},\text{odd})$ and $\bN$ belonging to an fcc
lattice. We Fourier transform according to 
\begin{equation}
\label{FT}
\phi^A_\bN (\tau)=\sqrt{\frac{2}{N_s}} \sqrt{\frac1{\beta}}
\sum_{\bk,\omega} \phi^A_k e^{i(\bk \cdot \bN-\omega \tau)} \times \left\{ \begin{array}{ll} 1 & A=\text{even} \\ e^{-ik_z/2} & A=\text{odd} \end{array} \right. .
\end{equation}
Here $k\equiv(\bk,\omega)$. In momentum space the action is given by
$S=S_2+S_3+S_4$ with (discarding {\it Umklapp} terms)
\begin{eqnarray}
S_2&=&\Tr \sum_k \left( \begin{array}{cc} \pi_k^{\e \dag} & \pi_{-k}^\o \end{array} \right)
\left( \begin{array}{cc} -i\omega + 2dJ & -2Jd \gamma_\bk \\
        -2Jd \gamma_\bk & i\omega + 2Jd \end{array} \right)
\left( \begin{array}{c} \pi_k^\e \\ \pi_{-k}^{\o \dag} \end{array} \right) \nonumber \\
&&\qquad \quad +\left( \begin{array}{cc} \chi_k^{\e \dag} & \chi_{-k}^\o \end{array} \right)
\left( \begin{array}{cc} -i\omega & 0  \label{S2} \\
        0 & i\omega \end{array} \right)
\left( \begin{array}{c} \chi_k^\e \\ \chi_{-k}^{\o \dag} \end{array} \right), \\
S_3&=&2Jd \sqrt{\frac2{N_s N_c \beta}} \Tr \sum_{kp} \left(
  \chi_p^{\o \dag} \chi_k^\e \pi_{p-k}^\o \gamma_\bk - \chi_k^{\e \dag}
  \chi_{k-p}^\o \pi_p^\e \gamma_{\bk-\bp} + h.c. \right), \label{S3} \\
S_4&=&2Jd \frac2{N_s N_c \beta}  \Tr \sum_{kpq}  \left(
    \chi_k^\e \chi_p^{\e \dag} \chi_q^\o \chi_{q+k-p}^{\o \dag}
    -\pi_k^\e \pi_p^{\e \dag} \pi_q^{\o \dag} \pi^\o_{q-k+p}
    -\pi_k^{\e \dag} \pi_p^\e \pi_q^\o \pi^{\o \dag}_{q-k+p} \right. \nonumber \\
&&\qquad \qquad \qquad \qquad \left. -\chi_k^{\e \dag} \chi_p^\e \pi_q^\o \pi_{q+k-p}^{\o \dag} -
  \pi_k^\e \pi_p^{\e \dag} \chi_q^{\o \dag} \chi_{q-k+p}^\o \right) \gamma_{\bk-\bp} \\
&&\qquad \qquad \qquad \qquad +\frac12 \left( \left( \pi_k^{\e \dag} \pi_p^\e + \chi_k^{\e \dag}
    \chi_p^\e \right) \left( \pi^\o_{k-p-q} \pi_q^\e + \pi_{-q}^{\e
      \dag} \pi_{-(k-p-q)}^{\o \dag} \right) \right. \nonumber \\
&&\qquad \qquad \qquad \qquad \qquad  \left. \left( \pi_k^{\o \dag} \pi_p^\o + \chi_k^{\o \dag} \chi_p^\o
    \right) \left( \pi^\e_{k-p-q} \pi_q^\o + \pi_{-q}^{\o \dag}
      \pi_{-(k-p-q)}^{\e \dag} \right) \right) \gamma_{\bk-\bp-\bq} \nonumber
 \label{S4} .
\end{eqnarray}
$\gamma_\bk$ is given by
\begin{equation}
\label{gamma}
\gamma_{\bk}=\frac1{d} \sum_{\mu=1}^d \cos{k_{\mu}/2}.
\end{equation}

At large $N_c$, $S_3$ and $S_4$ are small perturbations. The bare propagators can be read from $S_2$,
\begin{eqnarray}
\left( G^{\pi}_k \right)_{A B} &=& \frac1{\omega^2+4J^2 d^2 E^2_\bk} \left( \begin{array}{cc}
                i\omega + 2Jd & 2Jd \gamma_\bk  \\
                2Jd \gamma_\bk    & -i\omega + 2Jd \end{array}
            \right)_{A B} ,  \label{Gopi} \\
\left( G^{\chi}_k \right)_{A B} &=& \left( \begin{array}{cc}
                -1/i\omega & 0 \\
                0       & 1/i\omega \end{array} \right)_{A B}. \label{Gochi}  
\end{eqnarray}
(The propagators are diagonal in
the internal group indices.) Here $E_\bk=\sqrt{1-\gamma_\bk^2}$. The poles of the propagators give the
dispersion relations of the various bosons at $O(1)$. The result is
\begin{equation}
\omega^2+(2Jd E_\bk)^2=0 \qquad \qquad \text{for $\pi$},     \label{Eopi}
\end{equation}
and
\begin{equation}
\omega=0 \qquad \qquad \qquad \qquad \qquad \text{for $\chi$}. \label{Eochi}
\end{equation}

The $\pi$ excitations are $2(N-m)^2$ AF spin
waves. At $B=0$ they are the only excitations. It is easy to verify
that the Euclidean dispersion relation is $\omega^2 \sim -|\bk|^2$ at low momentum.
The $\chi$ excitations are $4(N-m)(2m-N)$ zero modes (``soft'' modes). Their energy is zero for
all momentum, a sign of the local degeneracy of the classical ground
state discussed above. 

We conclude this subsection by classifying the different excitations according to their $U(2m-N)\times U(N-m) \times U(N-m)$ representations. We denote a representation by
\begin{equation}
(r_1,r_2,r_3)^{(q_1,q_2,q_3)}.
\end{equation}
Here $r_1$ denotes the representation of $SU(2m-N)$, and $r_{2,3}$ denote the representations of the two $SU(N-m)$ subgroups. $q_i$ are the charges of the excitations under the remaining $U(1)$ factors of the unbroken subgroup. These are generated by the following diagonal matrices.
\begin{equation}
1_N, \qquad \left( \begin{array}{ccc} 1_{2m-N} & 0 & 0 \\
                                                0 & 0 & 0 \\ 
                                                0 & 0 & 0 \end{array} \right), \quad \text{and} \quad \left( \begin{array}{ccc} 0 & 0 & 0 \\
               0 & 1_{N-m} & 0 \\
               0 & 0 & -1_{N-m} \end{array} \right).
\end{equation}
We give the different representations in Table~\ref{reps}. The crucial point is that the AF spin waves and the zero modes reside is completely different representations. This means that the separation between these excitations in the $O(1)$ spectrum will survive at higher orders and that mixing can not occur.

\begin{table}[ht]
\begin{center}
\begin{tabular}{|c|c|c|} \hline
Field  & Representation & Dimension \\ \hline \hline
$\pi$ & $\left (1,N-m,\overline{N-m} \right)^{(0,0,+2)}$ & $(N-m)^2$\\ \hline
$\chi_{\text{even}}$ & $\left( 2m-N,1,\overline{N-m} \right)^{(0,+1,+1)}$ & $(2m-N)(N-m)$\\ \hline
$\chi_{\text{odd}}$ & $\left( 2m-N,\overline{N-m},1 \right)^{(0,+1,-1)}$ & $(2m-N)(N-m)$ \\ \hline
\end{tabular}
\caption{$U(2m-N)\times U(N-m) \times U(N-m)$ representations of the excitations. The conjugate fields belong to the conjugate representations. $n(\overline{n})$ stands for the fundamental(its conjugate) representation of $SU(n)$ and $1$ stands for a singlet. } \label{reps}
\end{center}
\end{table}
\subsection{Self energy calculation and $O(1/N_c)$ dispersion relations of the
  $\chi$ fields \label{sec:Self_E}}

\begin{figure}[ht]
\includegraphics[clip]{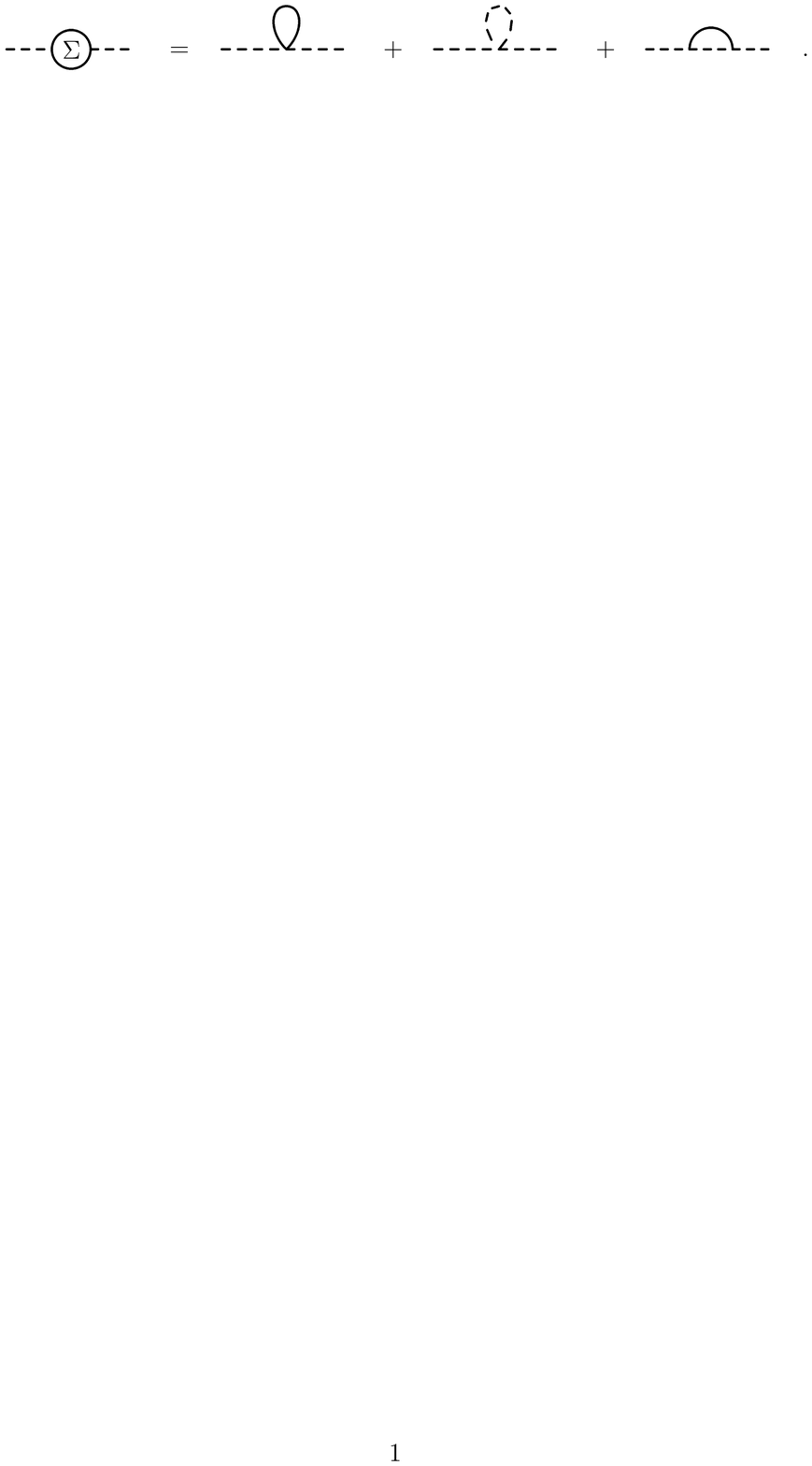}
\caption{Dyson equation for the self energy of the $\chi$ fields. Solid lines correspond to the $\pi$ propagators and
dashed lines to the $\chi$ propagators.} \label{fig:dyson}
\end{figure}

In this section we calculate the self energy of the $\chi$ fields to
first order in $1/N_c$. We shall see that the poles in the soft modes propagators
move away from zero energy in this order. The Feynman rules
for this problem are presented in Appendix~\ref{app:Feynman}. We treat $S_3$ and $S_4$ as
perturbations to $S_2$ and present the contributions to the self energy in Fig.~\ref{fig:dyson}.
  
Since the $\chi$ propagator is divergent at $\omega=0$ for all
$\bk$, one must insert its  self energy self-consistently, making the replacement
\begin{equation}
\left( G^{\chi}_k \right)^{-1} \rightarrow \left( G^{\chi}_k
\right)^{-1} - \Sigma_k. \nonumber
\end{equation}
$\Sigma$ is a matrix coupling odd and even degrees of freedom
according\footnote{We assume that the $\chi$ propagator has a similar
  structure as \Eq{Gopi}.} to
\begin{equation}
\Sigma=\left( \begin{array}{cc}
    \Sigma_1 & \Sigma_2 \\
    \Sigma_2^* & \Sigma_1^* \end{array} \right).
\end{equation}
Thus the $\chi$ propagator is replaced by
\begin{equation}
\label{Gchi}
G^{\chi}_k = \frac1{(-i\omega - \Sigma_{1,k})(i\omega - \Sigma_{1,k}^*)-|\Sigma_{2,k}|^2} \left( \begin{array}{cc}
        i\omega - \Sigma_{1,k}^* & \Sigma_{2,k} \\
        \Sigma_{2,k}^*   &       -i\omega - \Sigma_{1,k}
      \end{array} \right) .
\end{equation}
Here we assume that both matrices $\Sigma_{1,k}$ and $\Sigma_{2,k}$
are diagonal in group space and that
\begin{equation}
\label{stability}
\Re  \Sigma_{1,k}  < 0
\end{equation}
for the stability of the path integral.

In Appendix~\ref{app:Intequation} we derive self-consistent
equations for $\Sigma_{1,k}$ and $\Sigma_{2,k}$. Defining 
$\tanh{\theta_\bk}=-\Re \Sigma_{2,(0,\bk)} / \Re \Sigma_{1,(0,\bk)}$,
we write these
equations in the form of a single integral equation,
\begin{equation}
\label{Inteq}
(N-m)\tanh{\theta_\bk}=\frac{ {\displaystyle \int_{\text{BZ}} \left( \frac{dq}{4\pi}
  \right)^d I_2(\bq,\bk) \sinh{\theta_\bq} } }{ {\displaystyle \int_{\text{BZ}} \left(
    \frac{dq}{4\pi} \right)^d 
  I_1(\bq,\bk) \cosh{\theta_\bq} -\eta(\bk) } } \equiv \frac{\beta_\bk}{\alpha_\bk}.
\end{equation}
$\eta(\bk)$ and $I_{1,2}(\bq,\bk)$ are defined in Appendix~\ref{app:Intequation}. The
poles of the propagators may then be obtained from~(\ref{Gchi}) [see
\Eq{Eq4poles}]. The dispersion relations turn out to be
\begin{eqnarray}
\pm i\omega &=&\frac{2Jd}{N_c}\sqrt{(N-m)^2\alpha^2_\bk-\beta^2_\bk}
\nonumber \\
&=&\frac{2Jd(N-m)}{N_c}\alpha_\bk \sqrt{1-\tanh^2{\theta_\bk}}
 \\
&\equiv& \frac{2Jd(N-m)}{N_c} \epsilon_\bk. \nonumber
\end{eqnarray} 

\begin{figure}[htb]
\resizebox{100mm}{!}{\includegraphics{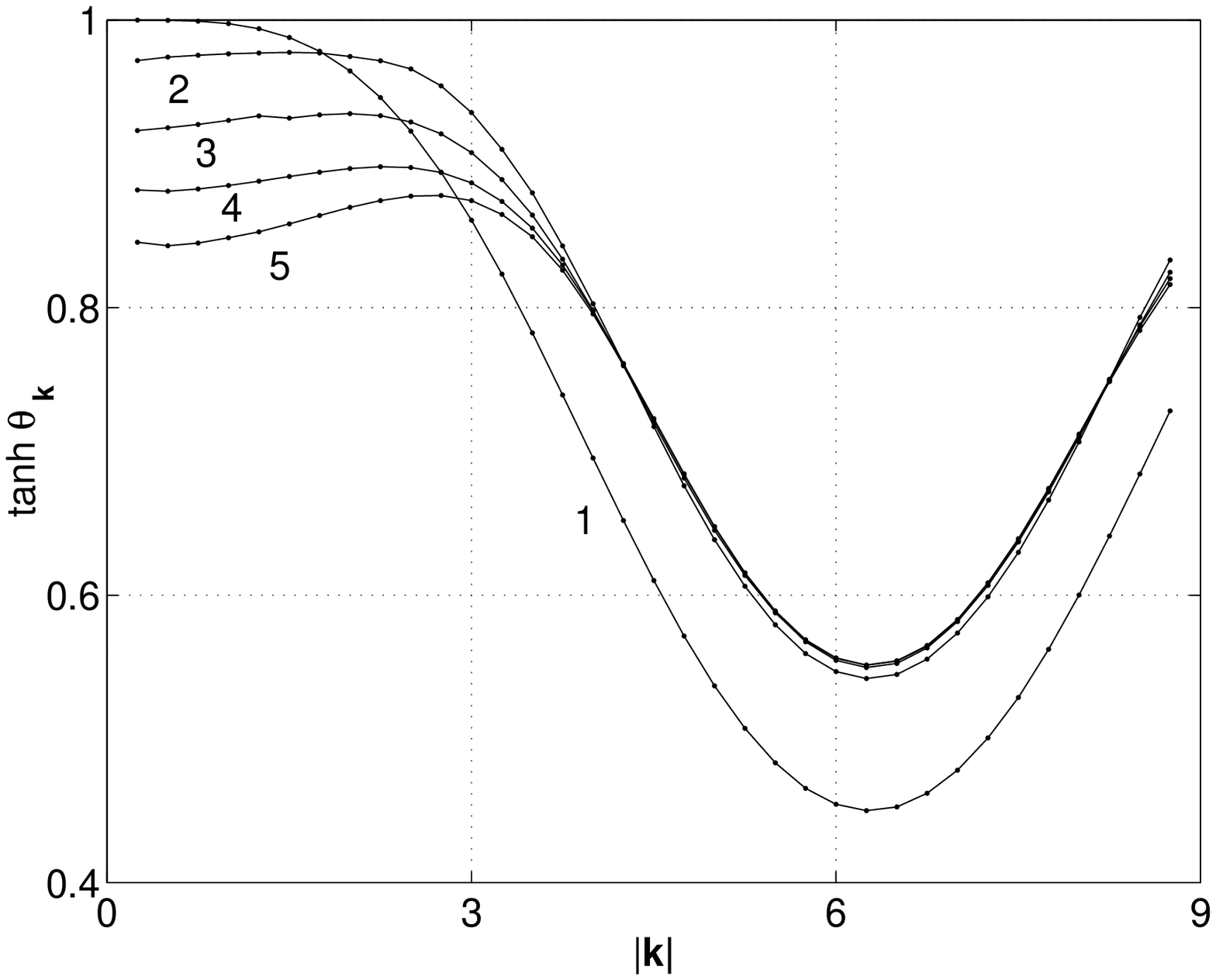}}
\caption{Solution of \Eq{Inteq} for $N-m=1,2,3,4,5$.} \label{fig:tanh}
\end{figure}

The solution of \Eq{Inteq} for $d=3$ can be obtained numerically by assuming
that the function $\theta_\bk$ depends only on $|\bk|$ in most of the 
 Brillouin zone. We plot the solution in
 Fig.~\ref{fig:tanh} for $N-m=1,2,3,4,5$. Recall that the baryon
 density is given by $B=m-N/2$. Thus for $N_f \le 3$ we cover all
 baryon density short of saturation.

\begin{figure}[htb]
\resizebox{100mm}{!}{\includegraphics{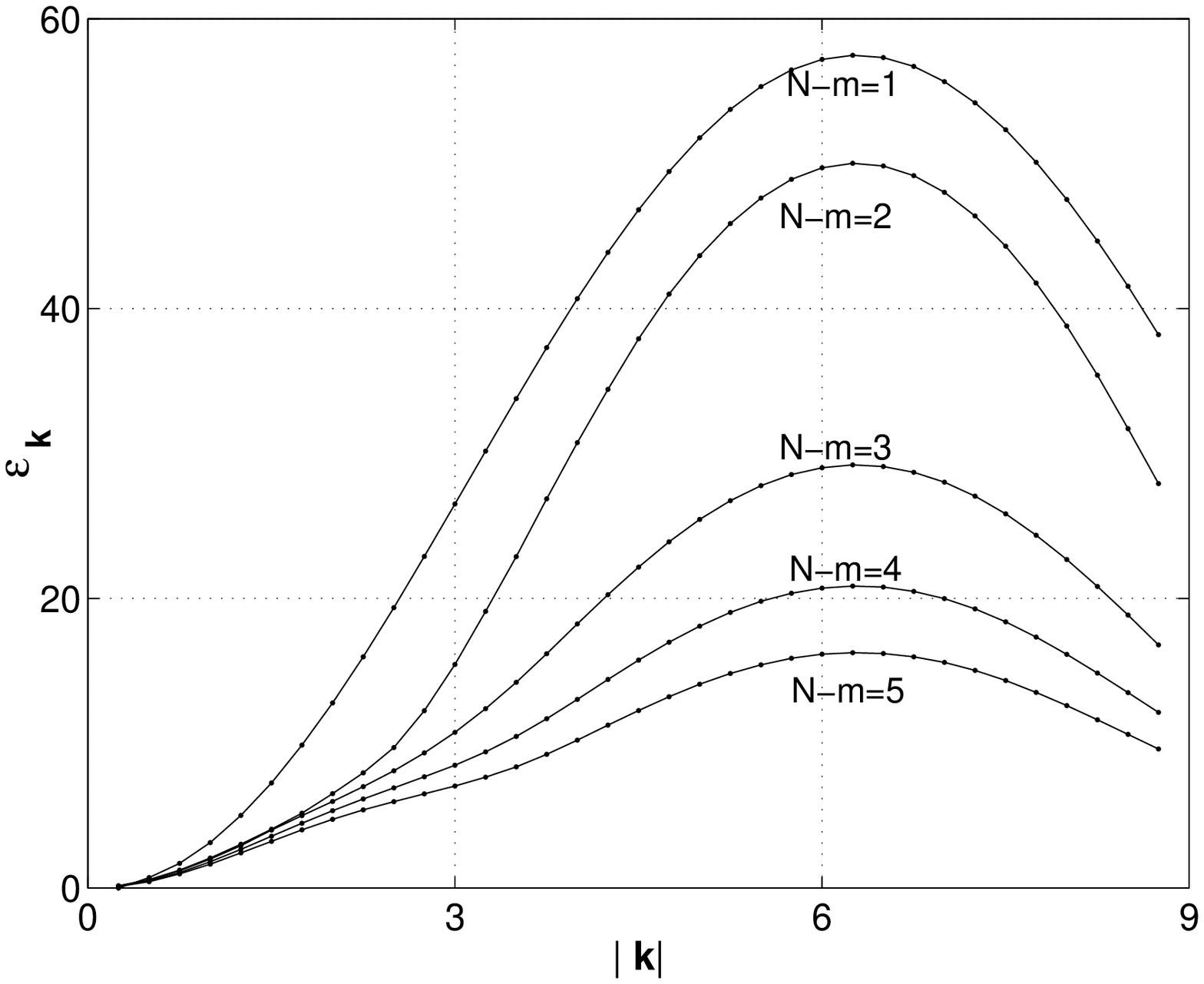}}
\caption{Rescaled energy of the $\chi$ bosons for 
  for $N-m=1,2,3,4,5$. The energy for $N-m=1$ is divided by a factor
  of $7$ in order to present it on the same scale with the other cases.} \label{fig:Ek}
\end{figure}

The form of $\epsilon_\bk$ is shown in Fig.~\ref{fig:Ek}.
It is easy to check that the bosons are massless, since
$\eta(0)=I_{1,2}(\bq,0)=0$ implies that $\alpha_{\bk}$ and $\beta_{\bk}$ 
vanish  at $\bk=0$. This is of course a direct result of the Ward identities
concerning the global $U(N)$ symmetry. Moreover, near $\bk=0$ both $\alpha$ and
$\beta$ have a quadratic dependence on $|\bk|$. This means that at low
momenta $\epsilon_\bk \sim |\bk|^{2}$, characteristic of
ferromagnetic magnons. A possible exception is the case $N-m=1$. There we see that at low momenta, $\tanh{\theta_\bk}\rightarrow
1$ and thus it is possible that $\epsilon_\bk \sim |\bk|^{2+p}$ with
$p$ a positive integer\footnote{$\epsilon_\bk$ cannot be nonanalytic
  at $\bk=0$ since the integrals in \Eq{Inteq} are regular there.}.

Finally we recall that the relation between the number of $\chi$
fields and
the number of physical excitations depends on the dispersion
relation. Because the soft modes obey $\epsilon_\bk \sim |\bk|^2$, the
$4(N-m)(2m-N)$ fields describe only half that many ferromagnetic magnons. The case of $N-m=1$ might be different
depending on whether $p$ is odd or even.

\section{Summary and discussion \label{sec:summary}}

We find two kinds of low energy excitations corresponding to type
I and type II Goldstone bosons \cite{Nielsen:hm,Leutwyler:1993gf}. The type I
bosons are $2(2N_f-B)^2$ antiferromagnetic spin
waves that have a linear dispersion relation. Including quantum fluctuations of $O(1/N_c)$ we also have  
$4B(2N_f-B)$ type II bosons\footnote{Recall that the local baryon number $B$ ranges from $0$ to $2N_f$ with unit increments.}. We call them ferromagnetic magnons since
they derive their energy from the effective ferromagnetic interaction \cite{Bringoltz:2002qc} between
next-nearest-neighbor sites, produced at $O(1/N_c)$. As typical for magnons, their dispersion relation is 
quadratic is momentum\footnote{A possible exception
  is for $N-m=1$, where the energy may depend on a different power of
  the momentum. Correspondingly, the number of excitations will be
  different.}.

The mechanism that removed the classical degeneracy and gave energy to the type II Goldstone bosons is known in the
 condensed
matter literature as order from disorder
\cite{Aharony,Ord_disorder,double_Xchange,Kagome,Sachdev,Henley}. It takes place in systems that posses a classical 
degenerate ground state. For example we mention the double exchange model
\cite{double_Xchange} and the Kagom{\' e} antiferromagnet
\cite{Kagome}. There, the classical ground state 
energy is invariant under a 
rotation of local groups of spins making it exponentially degenerate, like in our problem. It is worth recalling that the zero density system is also classically degenerate
 \cite{Bringoltz:2002qc,Smit:1980nf}.  
There, one can realize the baryon distribution by putting baryon
numbers $B$ and
$-B$ on the even and odd sites. At $O(1)$, the ground state energy does
not depend on $B$. This leads to a discretely degenerate ground state.
 At $O(1/N_c)$, order from disorder occurs and quantum fluctuation remove this degeneracy 
to pick out a ground state with $B=0$.

The energy scale of the type II Goldstone bosons is smaller by a factor of $N_c$ compared to that of the type I bosons.
 This points to a possible hierarchy of phase transitions at
finite temperature which can be described by a classical model that has the Hamiltonian
\begin{equation}
\label{Hmodel}
H=J_1 \hspace{0.1cm} \Tr \sum_{\bn \muhat} \left[ \sigma_\bn \sigma_{\bn +\muhat}
\right] - J_2 \hspace{0.1cm} \Tr \sum_{\bn \muhat} \left[ \sigma_\bn \sigma_{\bn+2\muhat} \right].
\end{equation}

For $J_2=0$, the ground state
will be highly degenerate as above. A small positive value of $J_2$ removes this
degeneracy and picks out a ferromagnetic alignment of the next-nearest-neighbor $\sigma$ fields. Thus for $T \ll J_{1,2}$ the ground state is 
invariant only under
\begin{equation}
U(N-m)\times U(N-m) \times U(2m-N).
\end{equation}
This symmetry pattern will persists until some finite temperature $T_c^{\text{FM}} \sim J_2$.
For $T > T_c^{\text{FM}}$ the ferromagnetic magnons melt the
magnetization and restore some of the symmetry.
This situation will persists until a second temperature $T_c^{\text{AFM}} \sim J_1$ where the symmetry will be restored completely. As long
as $J_1 \gg J_2$ (which corresponds to $N_c \gg 1$ in our system), the hierarchy of phase transitions is well defined.

Finally we mention that recent works \cite{Continuum_mu} on effective field theories for dense QCD also predict the
existence of type II Goldstone bosons. It is tempting to identify these with our ferromagnetic magnons. First, however, 
one must reduce the artificial $U(4N_f)$ 
symmetry of the action to the physical chiral symmetry. We defer this important issue to further publications.

\appendix
\section{Feynman rules \label{app:Feynman}}
\label{sec:App1}

\begin{figure}[ht]
\includegraphics[clip]{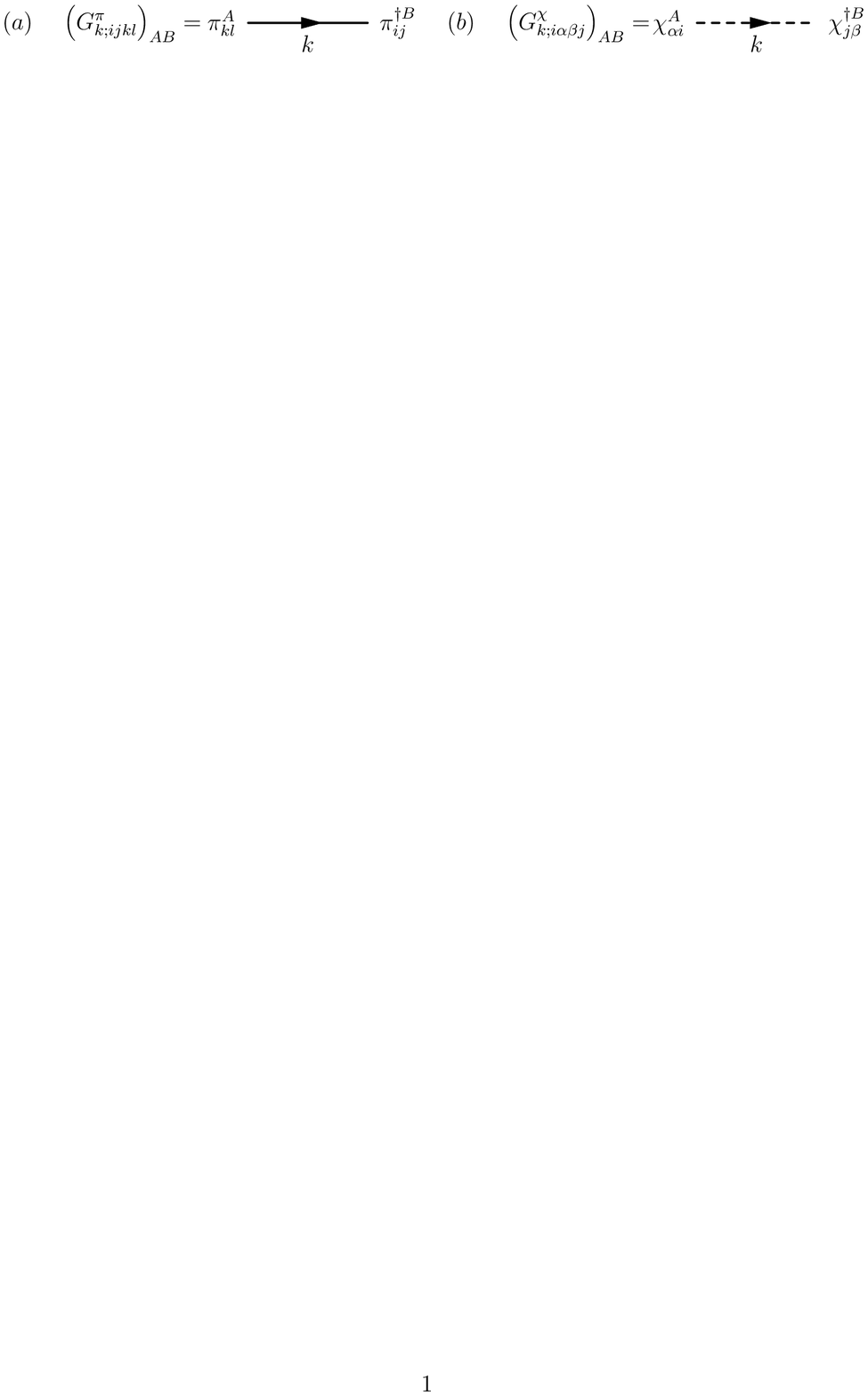}
\caption{Propagators.}
\label{fig:G}
\end{figure}

\begin{figure}[ht]
\includegraphics[clip]{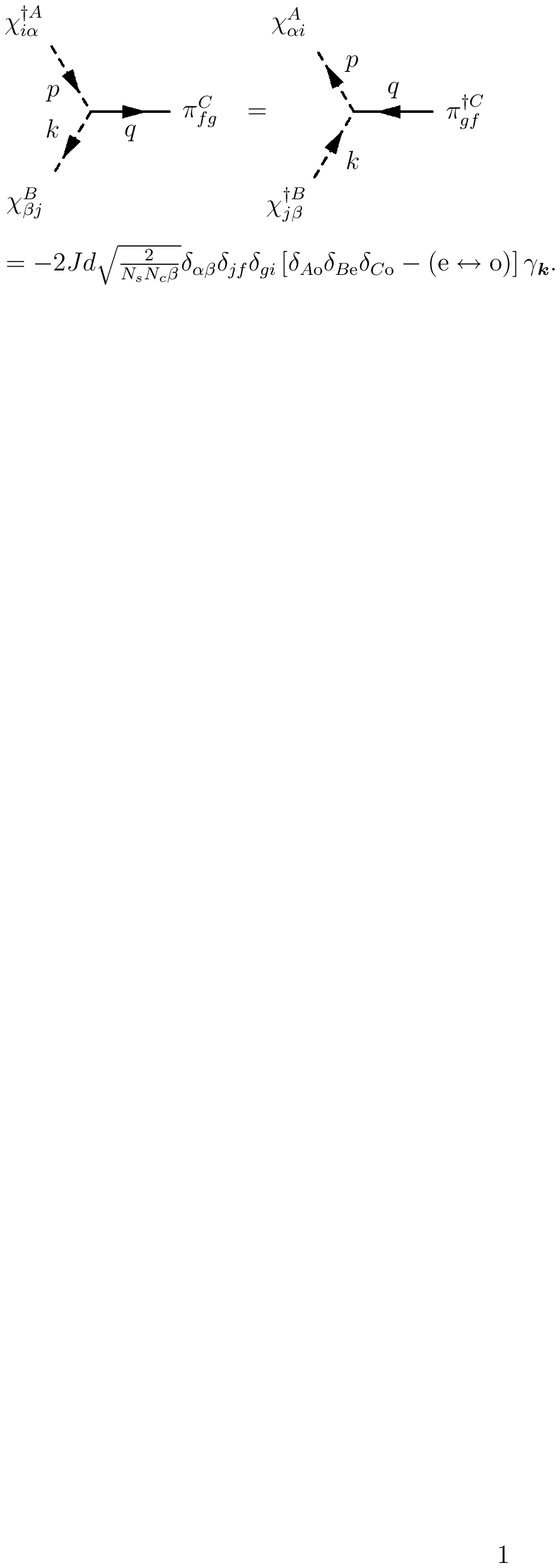}
\caption{Vertices corresponding to the cubic interaction.}
\label{fig:V3}
\end{figure}

The following are the Feynman rules extracted
from Eqs.~(\ref{S2})--(\ref{S4}). $A=(\e, \o)$ stands for even and odd
respectively. Latin indices take values in the range $[1,(N-m)^2]$ while Greek indices take values in the
range $[1,(N-m)(2m-N)]$. The propagators are given in Fig.~\ref{fig:G}. The vertices extracted from the cubic
 and quartic interactions [Eqs.~(\ref{S3})--(\ref{S4})] are given in Figs.~\ref{fig:V3}--~6.
\begin{figure}[ht]
\resizebox{!}{220mm}{\includegraphics[clip]{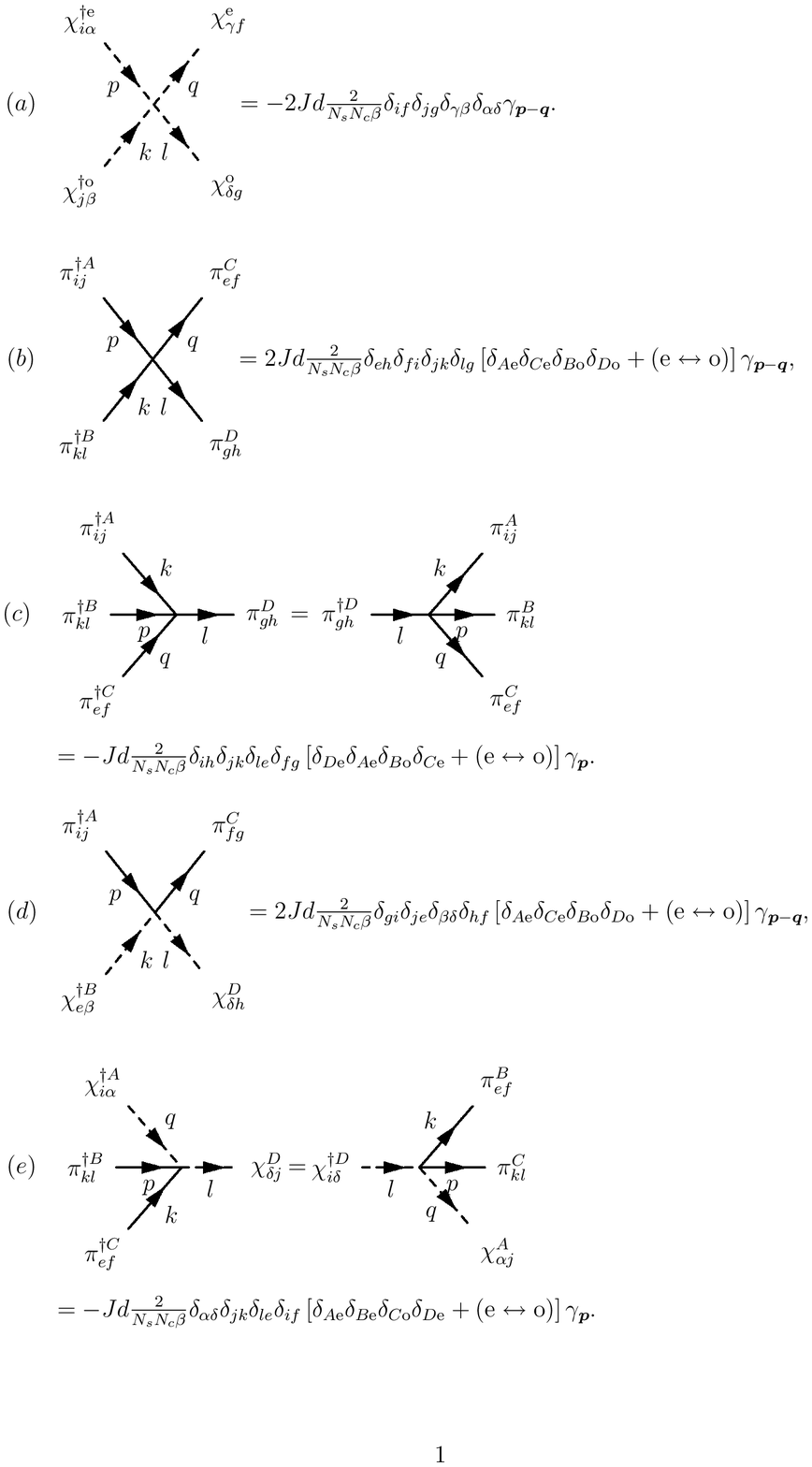}}
\label{fig:V4}
\caption{Vertices corresponding to the quartic interaction.}
\end{figure}

\section{The self consistent equations \label{app:Intequation}}
\label{sec:App2}
In this appendix we derive the integral equations represented by Fig.~\ref{fig:dyson} and solve them approximately in order to obtain the dispersion relations of the $\chi$ bosons to
$O(1/N_c)$.
In analogy with the propagators of the antiferromagnetic spin waves we
assume the following structure for $\Sigma_{1,k}$ and $\Sigma_{2,k}$ (we denote the frequency by $\Omega$), 
\begin{eqnarray}
\Sigma_{1,k}&=&A_1(\Omega^2;\bk)+i\Omega B_1(\Omega^2;\bk),  \label{Sig1} \\
\Sigma_{2,k}&=&A_2(\Omega^2;\bk)+i\Omega B_2(\Omega^2;\bk) \label{Sig2} ,
\end{eqnarray}
where $A_{1,2},B_{1,2}$ are real and $A^2_1 \geq A^2_2$ for
stability. In order to perform frequency integrals we make a rational
ansatz. For example, 
\begin{eqnarray}
A_1&=&-\frac{2Jd}{N_c}
\frac{P_1(\Omega^2;\bk)}{Q_1(\Omega^2;\bk)} < 0,
\label{A1} \\
A_2&=&\frac{2Jd}{N_c}
\frac{P_2(\Omega^2;\bk)}{Q_2(\Omega^2;\bk)} \label{A2},
\end{eqnarray}
Where $P_{1,2}(\Omega^2;\bk)$ and $Q_{1,2}(\Omega^2;\bk)$
are polynomials of the same order in $\omega^2$. Thus we have 
\begin{equation}
\label{Gchi_AB}
(G^{\chi}_k)_{A B} = \frac1{\Omega^2[(1+B_1)^2-B^2_2]+A_1^2-A_2^2} \left( \begin{array}{cc}
        i\Omega(1+B_1) - A_1 & A_2 +i\Omega B_2 \\
        A_2-i\Omega B_2   &       -i\Omega(1+B_1) - A_1 \end{array}
    \right)_{A B}. \qquad
\end{equation}
Since $B_{1,2}\sim O(1/N_c)$ we drop them in the denominator. The
self-consistent equations for $A$ and $B$ then decouple, and only the
former are needed to find the poles of the $\chi$ propagators. The equations\footnote{We have rescaled $\omega \rightarrow 2Jd \omega$ and
  $\Omega \rightarrow 2Jd \Omega$.} are
\begin{eqnarray}
\frac{P_1(\Omega^2;\bk)}{Q_1(\Omega^2;\bk)}&=&2(N-m) \int_{\text{BZ}} \left( \frac{dq}{4\pi} \right)^d \int_{-\infty}^{\infty} \frac{d\omega}{2\pi} \left[ -\frac{E^2_\bq}{(\omega-\Omega)^2+E^2_\bq} + \frac1{N_c}\frac{P_1(\omega^2;\bq)/Q_1(\omega^2;\bq)}{\omega^2+\tilde{E}^2(\omega^2;\bq)} \right. \nonumber \\
&& \left. -\frac{\omega(\omega-\Omega) \left(
      \gamma^2_\bq-\gamma^2_\bk \right)
    +\frac1{N_c}P_1(\omega^2;\bq)/Q_1(\omega^2;\bq) \left(
      2\gamma_\bk \gamma_\bq \gamma_{\bq-\bk} -
      \gamma^2_\bq-\gamma^2_\bk \right) }{\left[
      \omega^2+\tilde{E}^2(\omega^2;\bq) \right]  \left[
      (\omega-\Omega)^2+E^2_{\bq-\bk} \right] } \right],
\qquad \label{Ineq1_wq}
\end{eqnarray}
where the three terms come from the three diagrams in Fig.~\ref{fig:dyson} and
\begin{eqnarray}
\frac{P_2(\Omega^2;\bk)}{Q_2(\Omega^2;\bk)}&=&\frac2{N_c} \int_{\text{BZ}} \left( \frac{dq}{4\pi} \right)^d \int_{-\infty}^{\infty} \frac{d\omega}{2\pi} \frac{P_2(\omega^2;\bq)/Q_2(\omega^2;\bq)}{\omega^2+\tilde{E}^2(\omega^2;\bq)}
 \nonumber \\
 && \qquad \qquad \qquad \times \left[ \frac{2\gamma_\bk \gamma_\bq - \gamma_{\bq-\bk} \left(
      \gamma^2_\bq+\gamma^2_\bk \right)
    }{(\omega-\Omega)^2+E_{\bq-\bk}^2 } - \gamma_{\bq-\bk}
\right], \label{Ineq2_wq} \quad \qquad
\end{eqnarray}
which comes from the second and third diagrams in Fig.~\ref{fig:dyson}. Here
\begin{equation}
E_\bq=\sqrt{1-\gamma^2_\bq} \label{E}
\end{equation}
and
\begin{equation}
\tilde{E}(\bar{\omega}^2;\bq)=\frac1{N_c} \sqrt{
  \frac{P^2_1}{Q^2_1} - \frac{P^2_2}{Q^2_2} } \equiv \frac1{N_c}
\frac{P}{Q}. \label{E_tilde}
\end{equation}

It remains to evaluate the $O(1/N_c)$ contribution of the following integrals 
\begin{eqnarray}
I_1^i&=&\frac1{N_c}\int_{-\infty}^{\infty} \frac{d\omega}{2\pi}
\frac{P_i/Q_i}{\omega^2+\tilde{E}^2}, \label{I1} \\
I_2^i&=&\frac1{N_c}\int_{-\infty}^{\infty} \frac{d\omega}{2\pi}
\frac{P_i/Q_i}{\left( \omega^2+\tilde{E}^2 \right) \left[
    (\omega-\Omega)^2 + E^2 \right] }, \label{I2} \\
I_3&=&\int_{-\infty}^{\infty} \frac{d\omega}{2\pi} \frac{\omega ( \omega - \Omega) }{\left( \omega^2+\tilde{E}^2 \right) \left[ (\omega-\Omega)^2 + E^2
\right] } \label{I3} .
\end{eqnarray}
All these integrals are convergent at large $\omega$ and can be evaluated with residue calculus.
Beginning with $I_1^i$,
\begin{equation}
\label{I1cal0}
I_1^i=\frac1{N_c}\int_{-\infty}^{\infty} \frac{d\omega}{2\pi}
\frac{P_i Q_i Q_j^2}{\omega^2 Q^2 + \frac1{N_c^2}P^2}.
\end{equation}
The poles of the integrands are given by
\begin{equation}
\label{Eq4poles}
\omega^2 Q^2(\omega^2) + \frac1{N_c^2} P^2(\omega^2) = 0.
\end{equation}
To leading order in $1/N_c$ the roots are determined by either
\begin{equation}
\label{pole1}
\omega^2=-\frac1{N^2_c}P^2(0)/Q^2(0)
\end{equation}
or by
\begin{equation}
\label{more_poles}
Q^2(\omega^2)=0.
\end{equation}
Since the polynomials $P_i$ and $Q_i$ do not depend on $N_c$ and
$Q_i(0)$ can be chosen to be 1, the solutions of \Eq{more_poles} are
all $O(1)$. Moreover since $Q^2 > 0$, all the roots appear in complex
conjugate pairs $(\omega_n$, $\omega^*_n)$. For definiteness we choose
$\Im \omega_n >0$ and  close the contours from above. The leading
contribution to \Eq{I1} comes from the poles given by
\Eq{Eq4poles}. The calculation of $I_2$ and $I_3$ is similar, but one
has two more poles to consider, at $\omega=\Omega \pm iE$. The result
is 
\begin{eqnarray}
I_1^i&=&\frac12 \frac{P_i(0)}{\sqrt{P_1^2(0)-P_2^2(0)}}
\label{I1_result}, \\
I_2^i&=&\frac12 \frac{P_i(0)}{\sqrt{P_1^2(0)-P_2^2(0)}} \frac1{\Omega^2+E^2}, \label{I2_result} \\
I_3&=&\frac12 \frac{E}{\Omega^2+E^2} \label{I3_result}.
\end{eqnarray}
Thus we find that to $O(1/N_c)$, all integrals depend
only on the values of the polynomials at
$\omega=0$. Finally, Eqs.~(\ref{Ineq1_wq})--(\ref{Ineq2_wq}) simplify to
\begin{eqnarray}
\frac{P_1(\Omega^2,\bk)}{Q_1(\Omega^2,\bk)}&=&(N-m) \int_{\text{BZ}} \left( \frac{dq}{4\pi} \right)^d \left[ \cosh{\theta_\bq} \left( 1+\frac{2\gamma_\bk \gamma_\bq \gamma_{\bq-\bk} - \gamma^2_\bq - \gamma^2_\bk}{E^2_{\bq-\bk} +\Omega^2} \right)
\right. \nonumber \\
&& \left. -\frac{E_{\bq-\bk} \left( \gamma^2_\bq-\gamma^2_\bk \right)}{E^2_{\bq-\bk}+\Omega^2} - E_\bq \right],
\label{Inteq1_q} \\
\frac{P_2(\Omega^2,\bk)}{Q_2(\Omega^2,\bk)}&=& \int_{\text{BZ}}
\left( \frac{dq}{4\pi} \right)^d \left[ \sinh{\theta_\bq} \left(
    \frac{2\gamma_\bq \gamma_\bk-\gamma_{\bq-\bk} \left( \gamma^2_\bq
        + \gamma^2_\bk \right)}{E^2_{\bq-\bk}+\Omega^2} -
    \gamma_{\bq-\bk} \right) \right]. \label{Inteq2_q} \qquad
\end{eqnarray}
Where we have defined 
\begin{equation}
\label{tanh}
\tanh{\theta_\bq}=\frac{P_2(0,\bq)}{P_1(0,\bq)}.
\end{equation}

Taking $\Omega=0$, and dividing \Eq{Inteq2_q}
by \Eq{Inteq1_q}, we have 
\begin{equation}
(N-m)\tanh{\theta_\bk}=\frac{{\displaystyle \int_{\text{BZ}} \left( \frac{dq}{4\pi}
  \right)^d I_2(\bq,\bk) \sinh{\theta_\bq}  }}{{\displaystyle \int_{\text{BZ}} \left(
    \frac{dq}{4\pi} \right)^d I_1(\bq,\bk)\cosh{\theta_\bq} -\eta(\bk) }} \equiv \frac{\beta_\bk}{\alpha_\bk}.
\end{equation}
Here we have further defined 
\begin{eqnarray}
\eta(\bk)&=& \int_{\text{BZ}} \left( \frac{dq}{4\pi} \right)^d
\left[ \frac{\gamma^2_{\bq-\bk}-\gamma^2_\bk}{E_{\bq}} + E_\bq \right] \label{al0},       \\
I_1(\bq,\bk)&=& \left( 1+\frac{2\gamma_\bq \gamma_\bk \gamma_{\bq-\bk}-\gamma^2_\bq-\gamma^2_\bk}{E^2_{\bq-\bk}} \right), \label{a1}  \\
I_2(\bq,\bk)&=& \left(\frac{2\gamma_\bq
    \gamma_\bk-\gamma_{\bq-\bk}\left( \gamma^2_\bq+\gamma^2_\bk
    \right) }{E^2_{\bq-\bk}}-\gamma_{\bq-\bk} \right) \label{a2}.
\end{eqnarray}
A solution of this equation must obey
\begin{equation}
\label{stability1}
\int_{\text{BZ}} \left( \frac{dq}{4\pi} \right)^d I_1(\bq,\bk)
\cosh{\theta_\bq} -\eta(\bk) \geq 0,
\end{equation}
so that the stability condition~(\ref{stability}) will be satisfied.

After solving
 the integral equation~(\ref{Inteq}) one can extract the
spectrum of the $\chi$ bosons by calculating the poles of the
propagator~(\ref{Gchi}). These poles are given by
equation~(\ref{Eq4poles}). As seen there are two kinds of
solutions. The $O(1/N_c)$ poles of the form~(\ref{pole1}) give rise to
a low energy band [see \Eq{E_tilde}]
\begin{equation}
\label{E1k}
\pm i
\omega=\frac{2Jd}{N_c}\sqrt{P_1^2(0,\bk)-P_2^2(0,\bk)}
\end{equation}
Using \Eq{tanh} and \Eq{Inteq} we have 
\begin{equation}
\pm
i\omega=\frac{2Jd(N-m)}{N_c}\alpha_{\bk}\sqrt{1-\tanh^2{\theta_\bk}}
\equiv \frac{2Jd(N-m)}{N_c} \epsilon_\bk.
\end{equation}
The $O(1)$ poles that solve \Eq{more_poles} represent uninteresting
massive excitations.

\section{Hamiltonian approach for $N-m=1$  \label{app:Hamiltonian}}
\label{sec:App3}
In this appendix we derive the self-consistent equations in a
Hamiltonian formulation using a generalized Holstein-Primakoff
transformation for $N-m=1$.

The Hamiltonian is \cite{Bringoltz:2002qc} 
\begin{equation}
\label{Hqmapp}
H=\frac{J}2 \sum_{\bn \muhat \alpha \beta} Q_{\alpha \beta}(\bn)
Q_{\beta \alpha}(\bn+\muhat).
\end{equation}
The $U(N)$ generators $Q_{\alpha \beta}(\bn)$ obey
\begin{equation}
  \label{algebra}
  \left[ Q_{\alpha \beta}(\bn) , Q_{\gamma \delta}(\bmm) \right] =
  \left( Q_{\alpha
    \delta}(\bn) \delta_{\beta \gamma} - Q_{\gamma \beta}(\bn) \delta_{\alpha
    \delta} \right)\delta_{\bn \bmm} .
\end{equation}
For $N-m=1$ there is a simple representation of these operators
\cite{Randjbar-Daemi:1992zj},
\begin{equation}
  \label{realize}
  Q_{\alpha \beta} = \frac{N_c}2 \left( \begin{array}{ccc}
      {\bf 1}_{N-1}-2\phi \phi^{\dag} & & 2\phi\sqrt{1-\phi^{\dag} \phi} \vspace{0.3cm} \\
      2\sqrt{1-\phi^{\dag} \phi} \phi^{\dag} & & -1+2\phi^{\dag}\phi
    \end{array} \right)_{\alpha \beta}.
\end{equation}
Here $\phi$ is a $(N-1)$--component field that obeys 
\begin{equation}
  \label{commutations}
  \left[ \phi^{\dag}_i , \phi_j \right] = \frac1{N_c} \delta_{ij}.
\end{equation}
This representation is the generalized Holstein-Primakoff transformation. (We are not aware of such
 a representation for $N-m>1$. )
Next we write
\begin{equation}
  \label{phi_H}
  \phi=\left( \begin{array}{c} \chi \\ \pi \end{array} \right).
\end{equation}
The $(N-2)$--component field $\chi$ will represent the zero modes and the fields $\pi$ will constitute the antiferromagnetic spin waves. 

We write the generators on the even sites by inserting \Eq{phi_H} into \Eq{realize}. As in Section~\ref{sec:O1} we replace 
$Q$ by
\begin{equation}
Q'(\phi) \equiv V Q(\phi) V^{\dag}
\end{equation}
 on the odd sites. $V$ is given in \Eq{V}.
It is easy to see that the
  algebra~(\ref{algebra}) on the odd sites is obeyed by $Q'$ as well. 
  
Now, we scale $\phi \rightarrow
\phi/\sqrt{N_c}$ , substitute $Q$ and $Q'$ into \Eq{Hqmapp}, and 
  expand up to $O(1/N_c)$. We are left with
\begin{equation}
  \label{Heff}
H_{\text{eff}}=N_c^2 E_0 + N_c H_2 + \sqrt{N_c} H_3 + H_4, 
  \end{equation}
where
\begin{eqnarray}
  H_2&=&2J\left( 1-\frac{N-1}{N_c} \right) \sum_{\bn \muhat} \left(
    \pi_\bn \pi^{\dag}_\bn + \pi_{\bn+\muhat} \pi^{\dag}_{\bn+\muhat}
    -\pi_\bn \pi_{\bn+\muhat} -\pi^{\dag}_\bn \pi^{\dag}_{\bn+\muhat} \right),
    \label{H2} \\
  H_3&=& 2J\sum_{\bn \muhat,i} \left( \chi^e_{\bn i} \pi^o_{\bn+\muhat}
  \chi^{\dag o}_{\bn+\muhat,i} -
  \chi^e_{\bn i} \pi^{\dag e}_\bn \chi^{\dag o}_{\bn+\muhat,i} + h.c.
  \right), \label{H3} \\
  H_4&=& 2J\sum_{\bn \muhat} \left\{ \sum_{ij} \chi_{\bn i}
    \chi^{\dag}_{\bn j} \chi_{\bn+\muhat,j} \chi^{\dag}_{\bn+\muhat,i}
  \right. \nonumber \\
  && \qquad \quad \sum_i \left[ \frac12 \left( \pi_\bn \chi_{\bn i} \chi^{\dag}_{\bn i}
    \pi_{\bn+\muhat} + \pi_\bn \pi_{\bn+\muhat} \chi_{\bn+\muhat,i}
    \chi^{\dag}_{\bn+\muhat,i} + h.c. \right) \right. \nonumber \\
    &&\qquad \qquad \quad - \left( \left. \chi_{\bn i}\chi^{\dag}_{\bn i} \pi_{\bn+\muhat}
  \pi^{\dag}_{\bn+\muhat} + (\bn \leftrightarrow \bn+\muhat) \right) \right] \\
&& \qquad \quad \left. + \frac12 \left( \pi_\bn \pi_\bn \pi^{\dag}_\bn \pi_{\bn+\muhat} +
  h.c. \right) - \pi_\bn \pi^{\dag}_\bn \pi_{\bn+\muhat}
\pi^{\dag}_{\bn+\muhat} + (\bn \leftrightarrow \bn+\muhat) \right\} \nonumber
\label{H4}
\end{eqnarray} 

\subsection{The ground state at $O(1)$}

At lowest order in $1/N_c$, the Hamiltonian is given by $H_2$, which
does not depend on $\chi_\bn$. Moving to momentum space and
diagonalizing $H_2$ with a Bogoliubov transformation, we
have
\begin{equation}
  \label{H2q}
  N_c H_2=2Jd\left[ N_c-(N-1) \right] \sum_\bq \sqrt{1-\gamma^2_\bq} \left(
  a^{\dag}_\bq a_\bq + b^{\dag}_\bq b_\bq \right). 
\end{equation}
$\gamma_\bq$ is given by \Eq{gamma} and 
\begin{equation}
  \label{Bogoliubov}
  \left( \begin{array}{c} \pi^\e_\bq \\ \pi^{\dag \o}_{-\bq} \end{array}
  \right) = \left( \begin{array}{ccc} \cosh{\varphi_\bq} & &
      \sinh{\varphi_\bq} \\ \sinh{\varphi_\bq} & & \cosh{\varphi_\bq}
    \end{array} \right) \left( \begin{array}{c} a^{\dag}_\bq \\
      b_{-\bq} \end{array} \right).
\end{equation}
Here $\tanh{2\varphi_\bq}=\gamma_\bq$ and the fields $a_\bq$ and $b_\bq$ obey
the usual commutation relations of ladder operators.
The $O(1)$ ground state $|0 \rangle$ is thus given by
\begin{equation}
\label{Hgs}
a_\bq |0 \rangle = b_\bq |0 \rangle = 0.
\end{equation}
To $O(1)$ the excitations of $|0 \rangle$ are AF spin waves with a linear dispersion
relation. The ground state $|0 \rangle$ has a local degeneracy, corresponding to
arbitrary numbers of $\chi$ bosons (creation of a $\chi$ boson costs zero energy at this order). This is exactly the 
result of Section~\ref{sec:O1}.

\subsection{Self energy and $O(1/N_c)$ effective Hamiltonian}

In this section we calculate an effective Hamiltonian that splits
the spectrum of the degenerate subspace discussed above. We use Rayleigh-Schr\"{o}dinger perturbation
 theory \cite{Kato}.
$H_4$ contributes at first order and $H_3$
at second order. Both give a contribution of
$O(1/N_c)$. We must diagonalize the following Hamiltonian
within the degenerate subspace.
\begin{equation}
  \label{Hdeg}
  \tilde{H}_{\text{eff}}=P H_4 P+P H_3 \frac{1-P}{D} H_3 P.
\end{equation}
$P$ is a projection operator onto the degenerate sector and $D$ is the energy denominator. Once $\tilde{H}_{\text{eff}}$ is
calculated, we follow \cite{Aharony} and use the Wick theorem to
decouple its anharmonic terms in all possible
ways. This step includes substitution of various bilinears by their
vacuum expectation values. We calculate the vevs of the $\pi$
bilinears
using~(\ref{Bogoliubov}) and~(\ref{Hgs}),
\begin{eqnarray}
\langle \pi \pi^{\dag} \rangle^\e_\bn
&=&\langle \pi \pi^{\dag} \rangle^\o_{\bn \pm \muhat}
= \frac2{N_s} \sum_\bq \sinh^2{\varphi_\bq}, \label{vev1pi} \\
\langle \pi^{\dag \e}_\bn \pi^{\dag \o}_{\bn \pm \muhat} \rangle
&=&\langle \pi^\e_\bn \pi^\o_{\bn \pm \muhat} \rangle = 
\frac2{N_s} \sum_\bq \sinh{\varphi_\bq} \cosh{\varphi_\bq} e^{\pm i\bq
  \muhat}, \label{vev2pi}
\end{eqnarray}
and write an ansatz for the vevs of the $\chi$ bilinears,
\begin{eqnarray}
\langle \chi_i \chi^{\dag}_j \rangle^\e_\bn
&=&\langle \chi_i \chi^{\dag}_j \rangle^\o_{\bn \pm \muhat}
\equiv \frac1{N_s} \sum_\bq  \Delta_{1\bq} \delta_{ij},
\label{del1} \\
\langle \chi^{\dag \e}_{\bn j} \chi^{\dag \o}_{\bn \pm \muhat ,i} \rangle
&=&\langle \chi^e_{\bn i} \chi^o_{\bn \pm \muhat ,j} \rangle \equiv 
-\frac1{N_s} \sum_\bq \Delta_{2\bq} \delta_{ij}
e^{\pm i\bq \muhat}. \label{del2}
\end{eqnarray}
Here $\Delta_{1,2}$ are assumed to be real. 
Using Eqs.~(\ref{vev1pi})--(\ref{del2}) we decouple $H_4$
into the following form
\begin{equation}
\label{H4dec}
P H_4 P \simeq 2Jd \frac{2}{N_s} \sum_{\bq \bk ,i} \left( \begin{array}{cc}
    \chi^\e_{\bk i} & \chi^{\dag \o}_{-\bk i} \end{array} \right)
\left( \begin{array}{cc} \Delta_{1 \bq} +
    1-\sqrt{1-\gamma^2_q} & - \Delta_{2 \bq} \gamma_\bq
    \\
     - \Delta_{2 \bq} \gamma_\bq &  \Delta_{1 \bq}  +
    1-\sqrt{1-\gamma^2_q} \end{array} \right) \left(
  \begin{array}{c}
    \chi^{\dag \e}_{\bk i} \\ \chi^\o_{-\bk i} \end{array} \right). \qquad
\end{equation}

The calculation of of the second term in~(\ref{Hdeg}) is very
cumbersome. We will not pursue it here and just present the first
steps. We take the projection operator to be
\begin{equation}
\label{Q}
1-P=\sum_\bk |\bk_a \rangle \langle \bk_a| + |\bk_b \rangle
  \langle \bk_b |,
\end{equation}
where, for example $|\bk_a \rangle \equiv a^{\dag}_\bk |0
\rangle \otimes |\chi \rangle $. Here $ |\chi \rangle$ is $|\bk_a
\rangle$'s component in the $\chi$ Fock space. In general~(\ref{Q})  is not correct, since it takes into account
excitations of one spin wave only. 
Here it suffices since $H_3$ connects $|0 \rangle$ only with excitations of that sort. 
Next we replace the energy denominator with
\begin{equation}
\label{D}
D=2JdN_c \sqrt{1-\gamma^2_\bk} + O(1).
\end{equation}
This assumes that the energy of the $\chi$ bosons is smaller
than the energy of the spin waves by a factor of $N_c$.
After a lot of algebra we get
\begin{equation}
\label{Hresult}
\tilde{H}_{\text{eff}} \simeq 2Jd \sum_{\bk,i} \left( \begin{array}{cc}
    \chi^\e_{\bk i} & \chi^{\dag \o}_{-\bk i} \end{array} \right)
\left( \begin{array}{cc} J_{1 \bk} & J_{2 \bk} \\
    J_{2 \bk} &  J_{1 \bk} \end{array} \right) \left(
  \begin{array}{c}
    \chi^{\dag \e}_{\bk i} \\ \chi^\o_{-\bk i} \end{array} \right), \qquad
\end{equation}
with
\begin{eqnarray}
J_{1 \bk} &=& \int_{\text{BZ}} \left( \frac{dq}{4\pi} \right) ^d
\left( \Delta_{1 \bq} + 1 \right) I_1(\bq,\bk)
-\eta(\bk), \label{InteqH1} \\
J_{2 \bk} &=&  \int_{\text{BZ}} \left( \frac{dq}{4\pi} \right) ^d \Delta_{2 \bq} I_2(\bq,\bk). \label{InteqH2}
\end{eqnarray}
$I_{1,2}(\bq,\bk)$ and $\eta(\bk)$ are as
given in Eqs.~(\ref{a1})--(\ref{a2}). To get self-consistent equations we calculate the
vevs~(\ref{del1})--(\ref{del2}) using~(\ref{Hresult}). The
resulting equations are
\begin{eqnarray}
\Delta_{1 \bk}+1&=&\frac1{\sqrt{1-(J_{2\bk}/J_{1 \bk})^2}}, \label{eq1} \\
\Delta_{2 \bk}&=&\frac{J_{2 \bk}/J_{1 \bk}}{\sqrt{1-(J_{2\bk}/J_{1
      \bk})^2}}. \label{eq2}
\end{eqnarray}
Defining $\Delta_{1 \bk}+1 \equiv \cosh{\theta_\bk}$, $\Delta_{2
  \bk} \equiv \sinh{\theta_\bk}$ and dividing \Eq{eq2}
with \Eq{eq1} we get exactly
\Eq{Inteq} for $N-m=1$. 

This Hamiltonian calculation sheds light on the meaning of the
function $\theta_\bk$ in terms of vevs of the $\chi$ operators. A stable solution that obeys~(\ref{stability}) has
finite vevs. 
Mathematically, the integral
equation~(\ref{Inteq}) has also a trivial solution with $\theta_\bk=0$ that leads to 
$\Delta_{1,2}=0$.  However, this solution does not
obey~(\ref{stability}) and is thus unphysical, giving $J_1<0$ and making $\tilde{H}_{\text{eff}}$ not positive definite.

Finally, we recall that the absence of a Holstein--Primakoff transformation foretold
a Hamiltonian approach for $N-m>1$. The path integral approach is more general

\begin{acknowledgments}
I am indebted to B.~Svetitsky for helpful discussions and careful reading of the
manuscript. I also thank Amnon Aharony, Ora Entin-Wohlman and Dan Gl{\"u}ck 
for discussions. This work was supported by the Israel Science
Foundation under grant no.~222/02-1 and by the Tel Aviv University Research Fund.
\end{acknowledgments}

\end{document}